\documentclass[twocolumn,english,aps,pre,showpacs]{revtex4}
\usepackage{amsmath,amssymb,graphicx,bm,babel}

\begin{document}

\title{Effective surface dilatational viscosity of highly concentrated particle-laden interfaces}

\author{S.~V.~Lishchuk}

\affiliation{Department of Mathematics, University of Leicester, Leicester LE1 7RH, United Kingdom}


\begin{abstract}

The effective surface dilatational viscosity is calculated of a flat interface separating two
immiscible fluids laden with half-immersed monodisperse rigid spherical non-Brownian particles in
the limit of high particle concentration. The derivation is based upon the facts that (i)
highly-concentrated particle arrays in a plane form hexagonal structure, and (ii) the dominant
contribution to the viscous dissipation rate arises in the thin gaps between neighboring particles.

\end{abstract}

\maketitle


\section{Introduction}

Colloidal particles can adsorb irreversibly at the interfaces between two fluids \cite{Binks:CPLI}.
The capability of the adsorbed particles to stabilize emulsions has been known since the early 20th
century \cite{Ramsden:1903:156,Pickering:1907:2001}. This property has applications in many
industrial sectors, such as food processing \cite{Dickinson:2012:141}, biomedicine
\cite{Shilpi:2007:361}, petroleum industry \cite{Langevin:2004:511}.

Fluid interface with small concentration of the adsorbed particles can be regarded as
two-dimensional fluid \cite{Lishchuk:2009:56001}. It is possible to increase surface concentration
of particles by either adding new particles to the interface or changing surface area
\cite{Aveyard:2000:1969,Aveyard:2000:8820,Xu:2005:10016,Asekomhe:2005:1241,Monteux:2007:6344,Cheng:2009:4412}.
As the concentration increases, the interface laden with spherical particles undergoes transition to
2D crystal state \cite{Pieranski:1980:569} with possible intermediate hexatic phase between liquid
and solid phases \cite{Terao:1999:7157,Bonales:2011:3391,Qi:2014:5449}. Further increase in particle
concentration results in buckling of the interface
\cite{Aveyard:2000:1969,Aveyard:2000:8820,Xu:2005:10016,Asekomhe:2005:1241,Monteux:2007:6344,Cheng:2009:4412},
and allows producing stable non-spherical armoured bubbles \cite{Subramaniam:2005:930,Cui:2013:460}.

Highly concentrated particle-laden interfaces have recently received significant attention being the
basis of new soft materials with tunable properties. It has been proposed that colloidal
microcapsules formed by self-assembly of particles at the interface of emulsion droplets
(``colloidosomes'') can used as a tool for controlled delivery of drugs, food and cosmetic
supplements \cite{Dinsmore:2002:1006,Shilpi:2007:361,Dan:2012:141,Dommersnes:2013:2066}.
Non-sticking fluid droplets encapsulated with solid particles (``liquid marbles'') have potential
applications in sensors, chemical and biological microreactors, and droplet microfluidics
\cite{McHale:2011:5473,Bormashenko:2012:11018}. Bicontinuous interfacially jammed emulsion gels
(``bijels''), a new class of soft materials stabilized solely by colloidal particles, can
potentially serve as a cross-flow microreaction medium
\cite{Stratford:2005:2198,Herzig:2007:966,Cates:2008:2132}. Porous materials with controlled
porosity and pore sizes can be produced by removing fluid from particle-stabilized emulsions
\cite{Neirinck:2007:57}. The effectiveness of colloidal particles in stabilizing emulsions
(``Pickering emulsions'') also depends on the formation of a sufficiently dense layer of particles
at the fluid interface \cite{Tambe:1994:1}.

The surface rheology of particle-laden interfaces is an important factor in the stability of
particle-laden films, emulsions, and foams, as well as in their kinetics, such as break-up and
coalescence \cite{Fuller:2012:519,Mendoza:2014:303}. We may expect the surface rheology to play even
larger role at larger particle concentrations, when the gaps between particles are small but not
zero. Knowledge of interfacial rheology in this regime may be important for design of efficient
manufacturing procedure of the materials based on jammed particle-laden interfaces, mentioned above,
from the lower-concentration particle-laden systems, or for the prediction of their viscoelastic
properties. This motivates study of the rheology of highly concentrated particle-laden interfaces.

At sufficiently high concentrations the particle-laden interfaces exhibit viscoelastic behavior
\cite{Tambe:1994:1,Sagis:2011:1367}. Generally, viscous and elastic contributions to the surface
stress can be separated \cite{Verwijlen:2014:428}. Particle-laden interface at large scale can be
regarded as continuous, described, in particular, by the effective surface viscosities
\cite{Lishchuk:2009:016306}.

For isotropic interfaces the viscous contribution is well described by Boussinesq-Scriven model with
surface shear and dilatational viscosities as the material properties \cite{Scriven:1960:98}.
Isotropic change in surface area results in purely dilatational surface flow with the surface
velocity field
\begin{equation}
\label{eq:v_s}
\mathbf v_s=\alpha\mathbf r,
\end{equation}
where $\alpha$ is dilatation rate. Corresponding rate-of-strain tensor is isotropic,
\begin{equation}
\label{eq:S}
\bm{\mathsf S}=\alpha\bm{\mathsf I}_s,
\end{equation}
where $\bm{\mathsf I}_s$ is surface unit tensor. In this case viscous contribution to the surface stress
tensor, ${\bm\sigma_v=\zeta_s\bm{\mathsf S}}$, contains a single material parameter, dilatational
viscosity $\zeta_s$.

The origin of excess dissipation in particle laden interfaces lies in modification by the particles
of the flow in the bulk fluids that surround the interface. In the extreme case of small
concentration of the adsorbed particles the interaction between particles is small and the interface
is purely viscous. In this limit it is possible to calculate effective surface dilatational
viscosity of the fluid interface laden with half-immersed monodisperse spherical particles
\cite{Lishchuk:2009:016306}:
\begin{equation}
\label{eq:zeta-low}
\zeta_s=5(\eta_1+\eta_2)R\phi,
\end{equation}
where $\eta_1$ and $\eta_2$ are shear viscosities of the surrounding bulk fluids, $R$ is the radius
of the adsorbed particles, and 
\begin{equation}
\label{eq:phi}
\phi=\frac{\pi R^2N}A
\end{equation}
is the surface concentration of the particles, with $N$ being the number of particles in surface
area $A$.

In this work the result is presented of the calculation of the effective surface dilatational
viscosity in the opposite case of large particle concentrations, which complements low-concentration
result given by Eq.~(\ref{eq:zeta-low}).


\section{Derivation of the result}

\begin{figure}[tb]
\begin{center}
\includegraphics[width=\columnwidth]{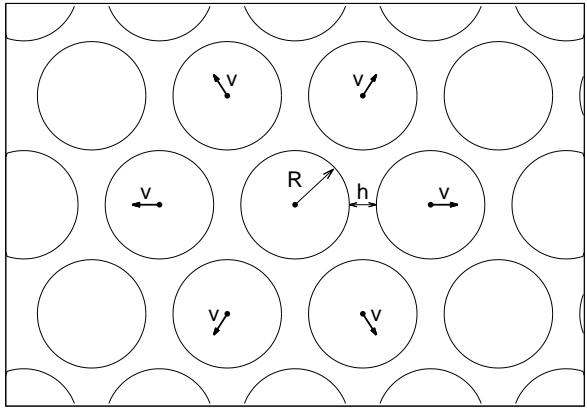}
\end{center}
\caption{\label{fig:geometry}
Hexagonal lattice of spherical particles under dilatational flow.
}
\end{figure}

We consider a system of identical rigid spherical particles of radius $R$ adsorbed at the flat
interface between two incompressible fluids. We neglect gravity and assume the interfacial tensions
favor a contact angle $\pi/2$, so that the particles are located with the equator coinciding with
the interfacial plane.

Different types of interaction between particles trapped at fluid interfaces are possible
\cite{Oettel:2008:1425}. We consider the case when the effective interaction between the particles
is repulsive, so that they do not agglomerate and the system remains homogeneous. The densest
packing of circles in the plane is the hexagonal (honeycomb) lattice, which has packing density
\begin{equation}
\label{eq:phi_m}
\phi_m=\frac\pi{2\sqrt3}
\end{equation}
We consider large surface concentration of the adsorbed particles, given by Eq.~(\ref{eq:phi}), and
assume that they are arranged in hexagonal lattice as shown in Figure~\ref{fig:geometry}. The gap
between the particles' surfaces, $h$, is expressed in terms of surface concentration of the
particles, $\phi$, by formula
\begin{equation}
\label{eq:h}
\frac1h=\frac1{2R}\left(\frac{\sqrt{\phi/\phi_m}}{1-\sqrt{\phi/\phi_m}}\right).
\end{equation}

We consider the system being subjected to the flow such that surface flow field, if unperturbed by
particles, would be purely dilatational, given by Eq.~(\ref{eq:v_s}). Due to symmetry of the system,
the neighboring particles move with respect to each other with relative velocity
\begin{equation}
\label{eq:v}
v=\alpha(2R+h),
\end{equation}
shown in Figure~\ref{fig:geometry}. We assume the interface to be macroscopically thin, having
surface tension sufficiently large to keep the fluid interface flat in the flow.

Frankel and Acrivos \cite{Frankel:1967:847} established that in highly concentrated
three-dimensional suspensions subjected to shear flow the dominant contribution to the viscous
dissipation rate comes from the gaps between neighboring particles. They found that the asymptotic
rate of viscous dissipation in the fluid with shear viscosity $\eta$ between two spheres approaching
each other with relative velocity $v$ is given by formula
\begin{equation}
\label{eq:E_3D}
\dot E=\frac{3\pi}4\eta v^2R\frac Rh+O\left(\ln\left(\frac hR\right)\right).
\end{equation}
The derivation was based on the exact solution of Stokes equations by Brenner
\cite{Brenner:1961:242}. Further terms in the expansion (\ref{eq:E_3D}) can be obtained using the
asymptotic expansion of lubrication force given by Jeffrey \cite{Jeffrey:1982:58}.

We consider the system with a flat particle-laden interface which is symmetric with respect to the
interfacial plane $z=0$. If shear viscosities of both bulk fluids are equal, the presence of the
interface between two bulk fluids does not change the solution of the hydrodynamic equations by
Brenner \cite{Brenner:1961:242} provided surface tangential stress is negligible. This condition is satisfied
if there is no additional adsorbed species at the interface, such as surfactants. In the case of
different viscosities there is a stress jump across the interface. It is straightforward to show
that tangential traction at the interface equals zero for the flow symmetric with respect to the
axis that joins the centers of two particles.
The normal traction due to pressure jump is compensated by surface tension if it is strong
enough, so that the interface remains flat. For this to occur, the Laplace pressure,
\begin{equation}
p_L\sim\frac\sigma R,
\end{equation}
$\sigma$ being surface tension, must be large compared to the pressure difference at the interface
between two fluids. In lubrication flow the excess pressure in the fluid is of order
$\eta vR/h^2$, yielding the pressure jump
\begin{equation}
p_\mathrm{flow}\sim\frac{|\eta_1-\eta_2||\alpha|R^2}{h^2}.
\end{equation}
The condition $p_L\gg p_\mathrm{flow}$ can be cast in form
\begin{equation}
\label{eq:alpha-condition}
|\alpha|\ll\frac{\sigma h^2}{|\eta_1-\eta_2|R^3}.
\end{equation}
meaning that interface is flat if dilatation rate $\alpha$ is small enough. This condition is
satisfied in most practical situations involving colloidal particles.

Thus, if the condition (\ref{eq:alpha-condition}) is satisfied there is no additional traction due
to presence of the interface between the fluids. As a result, the velocity field remains the same as
for the one-fluid problem without the interface. Viscous dissipation in the domain occupied by each
bulk fluid will be proportional to the value of corresponding shear viscosity. Therefore, the
asymptotic rate of viscous dissipation in the system can be written as a sum of the contributions
from each bulk fluid,
\begin{equation}
\label{eq:E_2D}
\dot E=\frac{3\pi}8(\eta_1+\eta_2)\frac{v^2R^2}h.
\end{equation}

In order to calculate the effective dilatational viscosity we shall follow the approach pioneered by
Einstein \cite{Einstein:1906:289} and equate expressions for the rate of viscous dissipation
calculated in two ways. First, we consider the system as homogeneous, having an effective continuum
interface with effective surface dilatational viscosity $\zeta_s$. Second, we consider the energy
dissipation in presence of particles explicitly.

The energy dissipation rate due to dilatational surface flow in a homogeneous surface characterized
by dilatational viscosity $\zeta_s$ is \cite{Scriven:1960:98}
\begin{equation}
\dot E_h=\zeta_s(\operatorname{Tr}\bm{\mathsf S})^2A,
\end{equation}
where the $A$ area of the interface. In our case $A$ equals the area of the hexagonal array of $N$
particles:
\begin{equation}
A=N\cdot\frac{\sqrt3}2(2R+h)^2.
\end{equation}
Substituting $(\operatorname{Tr}\bm{\mathsf S})^2=2\alpha^2$ in accordance with Eq.~(\ref{eq:S}), we
obtain
\begin{equation}
\label{eq:E_h}
\dot E_h=N\cdot\sqrt3\zeta_s\alpha^2(2R+h)^2.
\end{equation}

For dilatational flow in concentrated particle-laden interface we have
\begin{equation}
\dot E_p=3N\cdot\frac{3\pi}8(\eta_1+\eta_2)\frac{v^2R^2}h,
\end{equation}
where $3N$ is a number of lubrication areas between $N$ hexagonally arranged particles. Substituting
Eqs~(\ref{eq:phi_m}), (\ref{eq:h}) and (\ref{eq:v}) yields
\begin{equation}
\label{eq:E_p}
\dot E_p=N\cdot\frac{9\pi}{16}(\eta_1+\eta_2)\alpha^2R(2R+h)^2
\left(\frac{\sqrt{\phi/\phi_m}}{1-\sqrt{\phi/\phi_m}}\right).
\end{equation}

\begin{figure}[tb]
\begin{center}
\includegraphics[width=\columnwidth]{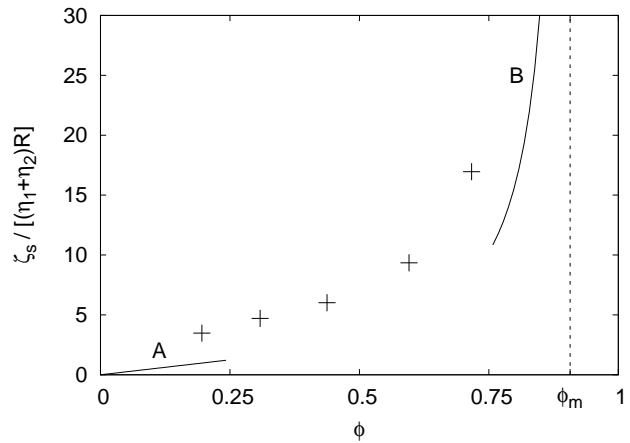}
\end{center}
\caption{\label{fig:zetas}
Asymptotic behavior of dilatational viscosity at low [line A, Eq.~(\ref{eq:zeta-low})] and high
[line B, Eq.~(\ref{eq:zeta-high})] particle concentrations. Crosses correspond to the numerical
results by Edwards and Wasan \cite{Edwards:1991:1247}.
}
\end{figure}

Equating two expressions for the energy dissipation rate, (\ref{eq:h}) and (\ref{eq:E_p}), we
finally obtain the formula for the effective dilatational viscosity at large concentration of
particles,
\begin{equation}
\label{eq:zeta-high}
\zeta_s=\frac{3\sqrt3\pi(\eta_1+\eta_2)R}{16\left(\sqrt{\phi_m/\phi}-1\right)}
\end{equation}
This result, together with the low-concentration asymptotics given by Eq.~(\ref{eq:zeta-low}), is
plotted in Figure~\ref{fig:zetas}.

\section{Discussion}

The derivation presented in the previous Section is based on the idea put forward by Frankel and
Acrivos \cite{Frankel:1967:847}, who suggested that fluid flow in the gaps between neighboring
particles gives the dominant contribution to the viscous dissipation rate in the highly concentrated
suspension, and used this idea to calculate the effective shear viscosity of highly concentrated
three-dimensional suspensions of solid spherical particles. We shall now discuss possible caveats of
this model and their implications on our result.

Frankel and Acrivos considered arrangement of particles in a lattice of fixed geometry. To
reconciliate their results with the experimental data, they employed the semiempirical cage model
\cite{Simha:1952:1020} to calculate viscous dissipation rate and found that the best agreement with
the available experimental data is achieved for the cubic arrangement of the particles. However, a
condition of fixed particle arrangement is not fulfilled in shear flow, and is at best an
instantaneous configuration. This fact together with non-uniqueness of the cage model makes the
result by Frankel and Acrivos empirical and requires fitting to the experimental data. Returning to
our system we note that, in contrast to shear flow, surface dilatational flow conserves hexagonal
arrangement of the particles on a surface, so there is no need in empirical parameters in our model.

Berlyand {\em et al} \cite{Berlyand:2009:585} considered fluid flow in highly concentrated strictly
two-dimensional suspensions and found that in two dimensions different types of singular terms
should be considered for adequate description of the divergence of the dissipation rate. In our
case this complication does not arise because, although particles are arranged on a 2D plane, fluid
flow between them remains three-dimensional. Moreover, of four types of flow considered in
Ref.~[\onlinecite{Berlyand:2009:585}], namely, the squeeze, the shear, and two types of rotation,
only the squeezing flow, considered in our derivation, is present in purely dilatational flow.

Marrucci and Denn \cite{Marrucci:1985:317} analyzed the result by Frankel and Acrivos and concluded
that for the realistic configurations of particles the averaging over relative pair positions of the
particles is incorrect, and the models based on regular lattice may present a misleading picture of
suspension behavior. In our case the deviation of particles' configuration from hexagonal lattice
also requires separate analysis. However, presence of the repulsive interaction between particles
sufficient to make them non-Brownian removes this, otherwise complicated, problem.

Edwards and Wasan \cite{Edwards:1991:1247} presented the results, shown in Figure~\ref{fig:zetas},
on the effective dilatational viscosity obtained by numerical solution of Stokes equations for the
case of adsorbed spherical particles arranged in a square lattice. It is not surprising that the
high-concentration results are higher than for hexagonal arrangement of particles because the
distance between particles arranged in a square lattice is less. It is less clear why the data at
lower concentrations is, being larger than predicted by low-concentration formula
(\ref{eq:zeta-low}), changes slower than proportionally to the particle concentration $\phi$.

Estimation of the concentration ranges at which the limiting formulas for the dilatational
viscosity, Eqs~(\ref{eq:zeta-low}) and (\ref{eq:zeta-high}), are accurate, requires separate
investigation. A simple criterium for the applicability of the high-concentration formula
(\ref{eq:zeta-high}) can be obtained by requiring viscous dissipation in lubrication gaps between
particles to be large compared to the rest excess dissipation in the surrounding fluids. This
requirement is equivalent to the condition of dilatational analog of Saffman-Delbr\"uck length
\cite{Saffman:1975:3111},
\begin{equation}
\label{eq:Saffman-Delbrueck}
l=\frac{\zeta_s}{\eta_1+\eta_2},
\end{equation}
being large compared to the characteristic length for the excess dissipation outside lubrication
gaps, which is of order of the particle size $R$. Expressing Eq.~(\ref{eq:zeta-high}) in terms of
$h$ and substitiuting the result in inequality $l\gg R$ yields
\begin{equation}
h\ll\frac{3\sqrt3\pi}8R\sim R,
\end{equation}
which is consistent with the initial assumption that the lubrication gap should be small compared to
the size of the particles.


\section{Conclusion}

We have obtained asymptotic expression for the effective surface dilatational viscosity of a flat
interface separating two immiscible fluids laden with half-immersed monodisperse rigid spherical
particles at high concentration of the particles. The derivation is based upon the facts that (i)
highly-concentrated particle arrays in a plane form hexagonal structure, and (ii) the dominant
contribution to the viscous dissipation rate arises in the thin gaps between neighboring particles.
Dilatational viscosity is given by Eq.~(\ref{eq:zeta-high}) and diverges as $h^{-1}$, where $h$ is
the distance between neighboring particles' surfaces.

It should be possible to verify this result by direct experimental measurements. In the case of
oscillatory flows the present result, obtained for the case of stationary flow, represents a
low-frequency limit of frequency dependent dilatational viscosity.

The result can be extended in several ways. It should be possible, with different degrees of
complexity, to extend the hydrodynamic model to the cases of different particle shapes, size
distributions, different contact angles between the particles and the fluid interface, different
curvatures of the interface, combined shear+dilatational flow ({\em eg} such as in Langmuir trough),
and so on. If particles adsorbed at the fluid interface are small (``nanoparticles''), physical
effects not captured by the simple hydrodynamic model become important and require further
investigation, for example, thermal motion of the particles and the interface
\cite{Bresme:2007:413101}, or modification of fluid viscosity due to confinement
\cite{Zhang:2004:10778,Hoang:2012:021202} and viscoelectric \cite{Dukhin:1993:1} effects.

The model for the dilatational viscosity can be used in conjunction with appropriate model for
interparticle interactions to describe viscoelasticity of particle-laden interfaces
\cite{Tambe:1994:1}. The present result should allow extending the description of viscoelastic
properties to the case of high concentration of the particles.

\acknowledgments

I thank Dr Ian Halliday and Dr Rammile Ettelaie for helpful discussions.




\end{document}